\documentclass{article}
\usepackage{amsmath}
\usepackage{psfig}

\newcommand{\bv}{ {\bf v} }

\newcommand{\br}{ {\bf r} }

\newcommand{\bq}{ {\bf q} }
\newcommand{\tq}{ \tilde q }
\newcommand{\hq}{ \hat q }
\newcommand{\btq}{ {\bf \tilde q} }
\newcommand{\bhq}{ {\bf \hat q} }

\newcommand{\bg}{ {\bf g} }

\newcommand{\bc}{ {\bf c} }

\newcommand{\btc}{ {\bf \tilde c} }
\newcommand{\tc}{  \tilde c }

\newcommand{\bU}{ {\bf U} }

\newcommand{\ttheta}{ \tilde \theta }

\newcommand{\fd}{\footnotesize{d}}
\newcommand{\dtot}[1]{ \frac{\mbox{\fd}}{\mbox{\fd} #1}}

\newcommand{\sech}{\text{sech}}

\begin{document}

\title{On Integrability and Chaos in Discrete Systems}
\author{
  Mark J. Ablowitz\\
  Department of Applied Mathematics\\
  University of Colorado-Boulder\\
  Boulder, Colorado, 80309, USA\\
  \\
  Yasuhiro Ohta \\
  Department of Applied Mathematics, \\
  Faculty of Engineering, Hiroshima University \\
  1-4-1 Kagamiyama, Higashi-Hiroshima 739, Japan\\
  \\
  A. David Trubatch\\
  Department of Applied Mathematics\\
  University of Colorado-Boulder\\
  Boulder, Colorado, 80309, USA\\ }

\date{}
\maketitle

\begin{abstract}
  The scalar nonlinear Schr\"odinger (NLS) equation and a suitable
  discretization are well known integrable systems which exhibit the
  phenomena of ``effective'' chaos. Vector generalizations of both the
  continuous and discrete system are discussed. Some attention is directed
  upon the issue of the integrability of a discrete version of the vector NLS
  equation.
\end{abstract}

\section{Introduction}

Integrable systems are usually thought to demonstrate regular temporal and
spatial evolution. Philosophically, it seems to be pleasing to separate
integrable motion from chaotic dynamics. The latter being the ``opposite'' of
the former. However, upon more careful investigation, there is not such a
simple distinction. Many well known integrable systems are actually ``on the
verge'' of exhibiting chaos whereby any perturbation: physical, or
numerical--e.g. even roundoff effects, can push them ``over the edge'' and
thereby excite chaotic
dynamics. In one dimension, the classical pendulum is such an example. And
in one space - one time dimension the well known integrable systems: the
sine-Gordon and nonlinear Schr\"odinger equations with periodic
boundary values are examples of integrable systems where complex and
irregular dynamics can be excited in certain regions of phase 
space \cite{AbCh95, AbHe96, AbSc94, AbSc93}.
The common feature in such situations is an initial value prescribed in the
proximity of special regimes of phase space, referred to as homoclinic
manifolds. Homoclinic manifolds are locations in phase space in which
highly sensitive solution structures exist. We have, in earlier papers,
discussed the notion of ``effective'' chaos  \cite{AbHe96,AbSc94}whereby any
perturbation can excite irregular and unstable motion--both of which are
manifestations of chaotic dynamics.

Chaotic behavior in perturbed integrable systems is significant for two
reasons: (i) Numerical schemes can be considered as perturbations of the
original system. If the perturbation of the numerical scheme introduces chaos,
the results of simulations will not reflect the dynamics of the unperturbed
system. (ii) In physical applications, perturbations --from
effects neglected in the derivation of the integrable equation--
can induce chaotic behavior behavior in the system. Even if this effective
chaos is only present  in certain regions of phase space, it is, in general,
impossible to determine {\em a priori} where these regions are.   

Finite-difference discretizations of integrable PDE's can be
considered both as perturbations of of the original PDE and as the basis for
numerical schemes. Here, we will consider spatial finite-difference
schemes. These finite-difference discretizations can be integrated
in time by well-known, highly-accurate methods for integrating ordinary
differential
equations. Numerical schemes based on finite differences are useful
because, unlike spectral schemes, finite-difference schemes
do not require special boundary conditions.

We begin by first discussing the scalar nonlinear Schr\"odinger
equation (NLS):

\begin{equation}
  i q_t = q_{xx} + 2q|q|^2. \label{nls}
\end{equation}

\noindent NLS is integrable via the inverse scattering transform
(IST) and has significant applications in physics, such as the evolution of
small amplitude slowly varying wave packets in: deep water, nonlinear optics
and plasma physics (see e.g. \cite{AbSe81}). 

Many discretizations of NLS have physical applications as discrete systems
(see e.g. \cite{CaBi95,EiLo85,KeCa86,KeTs87}). In particular, consider two
conservative (in fact, Hamiltonian)
finite-difference discretizations of NLS, the diagonal discrete NLS (DDNLS),

\begin{equation}
  i \dtot{t}q_n = \frac{1}{h^2} (q_{n-1} - 2 q_n +q_{n+1}) + 
    2 |q_n|^2 q_n   \label{ddnls}
\end{equation}

\noindent and the integrable discrete NLS (IDNLS),

\begin{equation}
  i \dtot{t}q_n = \frac{1}{h^2} (q_{n-1} - 2 q_n + q_{n+1}) + 
    |q_n|^2 (q_{n+1}+q_{n-1}).   \label{idnls}
\end{equation}

\noindent These two systems differ only in the
discretization of the nonlinear term, yet they have very different
properties. While both these discretizations retain the Hamiltonian
structure of the PDE, IDNLS (\ref{idnls}) is integrable via the
IST \cite{AbLa76b} while DDNLS is not. Differences in the dynamics of these
two systems
shed light on the role of integrability in the development of chaos in
perturbations. In fact, these two schemes can exhibit very different responses
to the same initial data. In the case of periodic boundary conditions, the
onset and structure of ``effective'' chaos is very different between
(\ref{ddnls}) and (\ref{idnls}) while, for the infinite lattice with a
rapidly-decaying boundary condition, (\ref{idnls}) has solitons
and (\ref{ddnls}) does not. The affect of non-conservative perturbations to
NLS has also been considered in the literature (e. g. \cite{McOv95}),
but we will not do so here.

The vector generalization of NLS,

\begin{equation}
  i \bq_t = \bq_{xx} + 2 \left\| \bq \right\|^2 \bq  \label{vnls}
\end{equation}

\noindent where $\bq$ is an $N$-component vector and $\left\| \cdot \right\|$
denotes the vector norm, can be integrated via the IST and has soliton
solutions \cite{Ma74} (actually, in \cite{Ma74} only the case $N=2$ was
studied in detail;
however the extension to the $N$-th order vector system is straightforward).

Vector NLS (eq. \ref{vnls}, hereafter referred to as VNLS) is applicable,
physically, under the similar
conditions as NLS (\ref{nls}) with the generalization that the field can have
more than one nontrivial component. For example, in optical fibers,
the two components of the second-order ($N=2$) VNLS correspond
to components of the electric field transverse to the direction of
wave propagation. These components of the transverse field
compose a basis of the polarization states (we note that, in general, a
non-integrable variation of VNLS is the appropriate model for optical fibers
\cite{Me87}, however, in certain circumstances (\ref{vnls}) is indeed the
appropriate equation \cite{Me89, EvMo92}).

Although certain perturbations of the integrable VNLS
have been investigated \cite{LaKa97,ShDo97}, the question of effective chaos
in VNLS has not been
studied. Discretizations of VNLS are natural candidates for
the study of chaos induced by perturbations. The research on effective chaos
for NLS suggests that the results of such studies will depend strongly on the
choice of discretization, particularly in the discretization of the nonlinear
term. Therefore, the choice of discretization for VNLS merits closer
analysis.

The system
\begin{equation}
  i \dtot{t} \bq_n =
    \frac{1}{h^2} (\bq_{n-1} - 2 \bq_n + \bq_{n+1}) +
    \left\| \bq_n \right\|^2 (\bq_{n-1}+\bq_{n-1})  \label{sydvnls}
\end{equation}
where $\bq_n$ is an $N$-component vector, manifests properties
if an integrable system (cf. Section \ref{section:symmetricsolitons}).
Moreover, this system \eqref{sydvnls} is extremely useful as a numerical
scheme for VNLS \eqref{vnls}. Therefore, the underlying chaotic nature of
some periodic solutions of \eqref{vnls} and \eqref{sydvnls} can be examined
effectively. Because \eqref{sydvnls} is a natural generalization of
IDNLS \eqref{idnls} that has some important symmetries of of VNLS , we refer
to this system as the symmetric discretization of VNLS or,
simply the symmetric system.

To date, no compatible linear operator scattering pair for the symmetric
system (\ref{sydvnls}) has been derived. Without such a pair (\ref{sydvnls})
cannot be solved via the IST. However,
under the reduction $\bq_n = e^{i \omega t} \bv_n$, where $\bv_n$ is
independent of $t$,  (\ref{sydvnls}) reduces to an
$N$-dimensional difference equation  which is known to be
integrable (cf. \cite{Su94}). Also, we note that the reduction
$\bq_n = \bc q_n$, where $\bc$ is a constant complex unit vector
($\| \bc \| = 1$) and $q_n$ a scalar, (\ref{sydvnls}) reduces to the
integrable discretization of NLS (\ref{idnls}). 

The symmetric discretization (\ref{sydvnls}) has a class of analytical
multi-soliton solutions \cite{AbTr98,Oh98}. These exact solutions of the
symmetric system only reduce to a subset of those associated with
(\ref{vnls}). In order to determine whether general multi-soliton
solutions exist, we examined, by numerical simulation, the interactions of
solitary waves
associated with the symmetric system which lie outside this class
of known analytical solutions.
The numerical evidence, discussed later in this paper, indicates that,
in the symmetric system, the solitary waves interact elastically and are
therefore true solitons. This  finding strongly suggests that the
symmetric system (\ref{sydvnls}) is indeed integrable.

In the future, we shall identify of regions of phase space where
effective chaos could be expected for the symmetric system. In these
regions of phase space, the dynamics of the symmetric discretization can be
compared, by numerical simulation, to the dynamics of different discretizations
of VNLS-- e.g. replacing the nonlinear term in (\ref{vnls}) by
$\| \bq_n \|^2 \bq_n$ in analogy with (\ref{ddnls}).

%
% Section: Effective Chaos in NLS
%

\section{Effective Chaos in NLS \label{chaos}}

Let us return briefly to the scalar equation (\ref{nls}).
In both discretizations (\ref{ddnls},\ref{idnls}), truncation error 
is $O(h^2)$. For the non-integrable  discretization (\ref{ddnls}) the
truncation error is a perturbation from integrability while the integrable
discretization (\ref{idnls}) is, as its name implies, integrable for finite
$h$. For NLS with periodic initial data, some
initial conditions evolve irregularly in space and time when numerically
integrated via DDNLS (\ref{ddnls}) at moderate discretizations.
At high levels of discretization this irregularity disappears though often
very slowly. This ``chaotic'' response is due to the perturbation from
integrability in the truncation error in DDNLS (\ref{ddnls}),
the non-integrable discretization. Simulations with IDNLS (\ref{idnls}) and
the same initial data do not show irregularity even at moderate levels of
discretization \cite{AbCh95, HeAb93}.

Irregular and chaotic dynamics can also be caused by
perturbations due to round-off error in the floating pont arithmetic of
computer simulation. This is demonstrated by integrating the initial data
with the integrable discretization (\ref{idnls}) via the {\em mathematically}
equivalent discretization

\begin{equation}
   i \dtot{t}\tq_n = \frac{1}{h^2} (\tq_{n-1} - 2 \tq_n + \tq_{n+1}) + 
    |\tq_n|^2 q_{n+1} + |\tq_n|^2 \tq_{n-1}.   \label{idnlst}
\end{equation}  

\noindent The only difference between the discretizations (\ref{idnls}) and
(\ref{idnlst}) is the distribution of the multiplication in the nonlinear term.
Nonetheless, in numerical simulations, the two solutions,
$q_n(t)$ and $\tq_n(t)$, with the same initial data ($q_n(0)=\tq_n(0)$)
can dramatically drift apart until their difference is comparable to the
amplitudes of the solution \cite{AbSc94,AbSc93}.

These instances of chaotic behavior in NLS are associated with proximity
of the initial data to homoclinic manifolds of NLS. The location and structure
of these homoclinic manifolds can be determined for
NLS by the spectral theory of the associated linear scattering problem.
When initial data are chosen well away from the homoclinic manifolds, the
chaotic responses described above are observed only at coarser
discretizations or not at all \cite{ AbCl91,AbSc93}. In low dimensional
systems, chaotic dynamics
in regions of phase space are associated with crossings of the homoclinic
manifold. However, by analyzing the spectral data of the associated linear
problem, one can demonstrate that for DDNLS, chaotic dynamics generically
occur without the solution crossing the homoclinic manifold \cite{AbHe96}.

%
%  Section: Vector NLS
%

\section{Vector NLS \label{pde}}

Solutions of VNLS (\ref{vnls}) are more complex than those of the scalar
case (\ref{nls}) because there are more degrees of freedom. In particular,
the dynamics of soliton interactions depends on the vector nature of the
system. A clear definition of solitons in the vector system is a prerequisite
for analyzing soliton solutions in the discretizations of VNLS such as 
the symmetric system (\ref{sydvnls}). Hence, here, we give a brief account of
some key properties of VNLS and their affect on the soliton solutions.

Under the reduction

\begin{equation}
  \bq = \bc q \label{redsol}
\end{equation}

\noindent where $\bc$ is a constant, $N$-component
vector such that $\left\| \bc\right\|^2 =1$ and $q$ is a scalar function of
$x$ and $t$, vector NLS (\ref{vnls}) reduces to NLS (\ref{nls}). 
This is a manifestation of the fact that the vector system (\ref{vnls}) 
is a generalization of NLS. When the vector-valued independent variable $\bq$
is restricted to a one-dimensional subspace
(which is equivalent to a scalar), NLS is recovered. As a consequence, any
solution of scalar NLS has a corresponding family of
solutions of VNLS. We call a solution of the form (\ref{redsol}) a
{\em reduction solution}. We refer to $\bc$ as the {\em polarization} of the
reduction solution.

The solitons of VNLS can be found by IST for the boundary condition of rapid
decay on the whole line \cite{Ma74}. The one-soliton
solutions of VNLS are the reduction solutions
(i.e. in the form of eq. \ref{redsol})  in which the scalar function $q$ is
a one-soliton solution of NLS and $\bc$ can be any unit vector.
In the backward ($t \rightarrow -\infty$) and
forward ($t \rightarrow +\infty$) long-time limits, a generic
rapidly-decaying whole-line solution of VNLS asymptotically approaches a
linear superposition of solitons:

\begin{equation*}
  \bq \sim \sum_j \bq_j^{\pm} \qquad \text{as} \quad
    t \rightarrow \pm \infty
\end{equation*}

\noindent where $\bq_j^{\pm} = \bc_j^{\pm} q_j^{\pm}$, $\bc_j^{\pm}$ is
a complex unit vector and $q_j^{\pm}$ is a one-soliton solution of NLS such
that $q_j^+$ has the same amplitude and speed as $q_j^-$ \cite{Ma74}. 
Therefore, in the long-time limits, we can define a separate polarization
vector for each  soliton, namely $\bc_j^{\pm}$. Comparison of the forward
($+$) and backward ($-$) long-time limits shows that, in general,

\begin{equation*}
  \bc_j^- \neq \bc_j^+.
\end{equation*}

\noindent The polarization of each individual soliton shifts due interaction
with the other solitons. However,

\begin{equation}
  ||\bc_j^+||^2 = ||\bc_j^-||^2 = 1. \label{vcons}
\end{equation}

\noindent In addition to the shift in the polarization, the 
locations of individual soliton peaks are shifted by the interaction with other
solitons (this is the usual phase shift seen in scalar soliton equations).
A closed formula for these phase shifts (the change in polarization
and in the location of the peak) can be derived via the IST \cite{Ma74}. 

We refer to the squared absolute value of a component of the polarization
vector  as the {\em intensity} of that component of the polarization.
The intensity of the $\ell$-th component of the $j$-th soliton is given by
$|c_j^{(\ell)+}|^2$  ($|c_j^{(\ell)-}|^2$) where $c_j^{(\ell)+}$
($c_j^{(\ell)-}$) is the $\ell$-th component of $\bc_j^+$ ($\bc_j^-$), the
polarization vector of  $j$-th soliton in the forward (backward)
long-time limit. The relation (\ref{vcons}) implies that, for each soliton,
the sum of the intensities is constant, namely equal to one. However, the
distribution
of the intensity among the components of each soliton will not, in general,
be equal in the forward and backwards long-time limits due to interaction
with other solitons: that is, subject to the constraint of eq. (\ref{vcons}),
in general,

\begin{equation}
  |c_j^{(\ell)-}|^2 \neq |c_j^{(\ell)+}|^2. \label{intshift} 
\end{equation}

\noindent This shift in the distribution of intensity between the backward
and the forward long-time limits (\ref{intshift}) is a distinctive feature
of soliton interactions in the vector system (\ref{vnls}). There is no
corresponding phenomenon in the scalar equation (\ref{nls}).

Manakov \cite{Ma74} showed by IST that the following special case
holds for $M$-soliton solutions: if, for any $j,k= 1\ldots M$, either 

\begin{subequations}
\begin{equation}
  |\bc_j^- \cdot \bc_k^-| = 1 \label{minus:parallel}
\end{equation}

\noindent or

\begin{equation}
  |\bc_j^- \cdot \bc_k^-| = 0, \label{minus:orthogonal}
\end{equation}
\end{subequations}

\noindent where $\cdot$ denotes the inner product, then it turns out that

\begin{equation}
  |\bc_j^- \cdot \bc_j^+ | = 1 \label{noshift}
\end{equation}

\noindent and, moreover,

\begin{equation}
 |c_j^{(\ell)-}|^2 = |c_j^{(\ell)+}|^2 \label{nointshift}
\end{equation}

\noindent for all $\ell=1,2$, $j=1, \ldots, M$ (we assume the case of two
components). These soliton solutions constitute a special class of vector
solitons where the distribution of intensity  of an individual soliton is
not shifted  (i.e. the relation (\ref{nointshift}) holds) even though the
solitons pass through each other.

When considered as a model for electromagnetic wave propagation in
optical fibers, the components of VNLS (\ref{vnls}) play the
role of a basis for the polarization vector of the transverse electric field.
Because this choice of basis is arbitrary, the vector system (\ref{vnls})
ought to be, and is, invariant under a change of basis. Mathematically, a
change of basis is obtained by the independent-variable transformation

\begin{equation}
  \bhq = \bU \bq \label{unitrans}
\end{equation}

\noindent
where $\bU$ is a unitary matrix. The new independent variable, $\bhq$,
satisfies VNLS if, and only if, $\bq$ does.  We note that, in particular,
the distribution of intensity for a soliton depends on the choice of
basis. However, the conditions (\ref{minus:parallel}-\ref{minus:orthogonal})
and the consequence (\ref{noshift}) are independent of the choice of
basis-- i. e. dot products are preserved under unitary transformations.

%
%  Section: Discretization of Vector NLS
%

\section{Discretization of Vector NLS}

As we noted previously, the symmetric system (\ref{sydvnls}) is the natural
vector generalization of IDNLS (\ref{idnls}). However, the most natural
vector generalization of the linear scattering pair for IDNLS yields the
following nonlinear equation as a result of compatibility:

\begin{subequations}
\begin{align}
  i \dtot{t} \bq_n &=
    \frac{1}{h^2} (\bq_{n-1} - 2 \bq_n + \bq_{n+1}) -
    (\br_n^T \bq_{n-1}) \bq_n - (\br_n^T \bq_n) \bq_{n+1}
    \label{asdvnlsq} \\
  -i \dtot{t} \br_n &=
    \frac{1}{h^2} (\br_{n-1} - 2 \br_n + \br_{n+1}) -
    (\br_n^T \bq_n) \br_{n-1} - (\br_{n+1}^T \bq_n ) \br_n
    \label{asdvnlsr}
\end{align}
\end{subequations}

\noindent where $\bq_n$ is an $N$-component vector as is
$\br_n$ and the superscript $T$ denotes the transpose. To get VNLS from
(\ref{asdvnlsq}-\ref{asdvnlsr}), we take $\br = -\bq^*$ 
($*$ denotes the complex conjugate) after taking
the continuum limit. The discrete system 
(\ref{asdvnlsq}-\ref{asdvnlsr}) does not admit the symmetry
$\br_n = -\bq_n^*$  for finite  $h$. Hence, we refer to
(\ref{asdvnlsq}-\ref{asdvnlsr}) as the asymmetric discretization of VNLS, or
more briefly as the asymmetric system.
We emphasize that (\ref{asdvnlsq}-\ref{asdvnlsr}) is not equivalent to the
symmetric system (\ref{sydvnls}).

The asymmetric system has a class of soliton solutions
which corresponds to all the soliton solutions of the continuous
version of VNLS (\ref{vnls}) \cite{AbTr98}. However, in these solutions,
$|\br_n - (- \bq_n^*)| = O(h)$ so the asymmetry cannot not be 
ignored \cite{AbTr98}. In general, for initial data such that
$\br_n = -\bq_n^*$, the solution will evolve so that $\br_n \neq -\bq_n^*$
at later times. The unavoidable asymmetry of the asymmetric system
(\ref{asdvnlsq}-\ref{asdvnlsr}) makes it less desirable as a discretization
of VNLS. Hence, we turn our attention to the symmetric discretization.

We now discuss some important properties of the symmetric discretization.
The symmetric system (\ref{sydvnls}) reduces to
IDNLS (\ref{idnls}) under the reduction

\begin{equation}
  \bq_n = \bc q_n \label{reduce}
\end{equation}

\noindent where $\bc$ is a constant vector such that $|| \bc || = 1$.
This is the discrete analog of eq. (\ref{redsol}). As for VNLS,
we refer to the vector $\bc$ in (\ref{reduce}) as the {\em polarization} of
the {\em reduction solution} $\bq_n$. Also, the symmetric system is invariant
under a discrete version of the independent-variable
transformation (\ref{unitrans}). Hence,
as for the PDE (\ref{vnls}), the components of $\bq_n$ can be considered
to be a basis of the polarization and the discrete form of the
unitary transformation (\ref{unitrans}) as a change of coordinates. Perhaps
the most important property of the symmetric system is the existence of soliton
solutions. These are discussed in the next section.

%
% Solitons of Symmetric DVNLS
%

\section{Soliton Solutions of the Symmetric System
  \label{section:symmetricsolitons}}

In order to demonstrate the existence of solitons in the symmetric system,
we must first identify the appropriate vector solitary waves. Given these
solitary wave solutions, we then investigate their interaction. To
construct a vector solitary-wave solution of the symmetric system
(\ref{sydvnls}) multiply the scalar one-soliton solution of
IDNLS (\ref{idnls}) by an arbitrary unit vector. The resulting
vector solution $\bq_n$ is the reduction solution-- i.e. of the form
(\ref{reduce}) --where the scalar function $q_n$ is given by:

\begin{equation}
  q_n(t) = A e^{- i ( b h n - \omega t - \phi)} \sech(a h n - v t - \theta)
         = e^{i \phi} \hq_n(t)
  \label{discsol} 
\end{equation}

\noindent with

\begin{equation*}
       A = \frac{\sinh(a h)}{h}, \qquad
  \omega = \frac{2 (1 - \cosh(a h) \cos (b h))}{h^2},  \qquad
       v = 2 \frac{\sinh(a h) \sin(b h)}{h^2}
\end{equation*}

\noindent where $a$,$b$,$\theta$,$\phi$ are arbitrary parameters.
For this vector solitary wave, $a$ and $b$ determine the amplitude,
$A$ and speed, $v$. The parameter $\theta$ is the position
of the peak of $||\bq||$ at $t=0$ and $\phi$ is an overall complex phase.

The simplest example of the interaction of these vector solitary waves is 
constructed from the scalar $M$-soliton solution of IDNLS. Let $q_{j,n}$
be the scalar $M$-soliton solution of IDNLS with parameters $a_j$, $b_j$
and consider the vector solution $\bq_n = \bc q_n$ where, as before,
$\bc$ is a unit vector. Then in the long-time limits
($t \rightarrow \pm \infty$),

\begin{equation}
  \bq_n^{\pm} \sim \bc \sum_{j=1}^{M} e^{i \phi_j^{\pm}} \hq_{j,n}^{\pm},
  \label{trivlim}
\end{equation} 

\noindent where $\hq_{j,n}^{\pm}$ is a solitary-wave solution of the
form given above in  eq. (\ref{discsol}) with $a=a_j$, $b=b_j$,
$\theta=\theta_j^\pm$. The parameters $\phi_j^{\pm}$ are complex phase shifts
of the scalar soliton
interaction. For such a solution, each solitary wave envelope can be thought
of as having an individual polarization:

\begin{equation*}
  \bc_j^{\pm} = \bc e^{i \phi_j^{\pm}}.
\end{equation*}

\noindent In this example, we see that each of the
vector solitary waves has the same amplitude and speed in the forward
long-time  limit as it had in the backward long-time limit (i.e. a
soliton-type interaction). This suggests that the individual vector solitary
waves constructed from (\ref{discsol}) are, in fact, the solitons of the
symmetric system (\ref{sydvnls}).

However, in this simple example (\ref{reduce}), since $\bc$ is the
same for each solitary wave, the polarizations of the vector solitary
waves must satisfy  

\begin{equation*}
  | \bc_j^- \cdot \bc_k^- | = 1    
\end{equation*}

\noindent for all $j,k = 1 \ldots M$. There is no such constraint
for the solitons in the vector PDE (\ref{vnls}). Therefore, we investigate
the more general vector solitary-wave interactions
(i.e. $| \bc_j^- \cdot \bc_k^- | < 1$) in the discrete symmetric system
by direct and numerical methods.

\subsection{Soliton solutions by Hirota's Method}

To find soliton solutions of the symmetric system (\ref{sydvnls}) by Hirota's
method we write it in the bilinear form: by taking

\begin{equation*}
   \bq_n = \frac{\bg_n}{f_n}.
\end{equation*}

\noindent where $\bg_n$ is a vector and $f_n$ is a scalar, we find

\begin{subequations}
\begin{gather}
   i h^2D_t f_n \cdot \bg_n = f_{n-1} \bg_{n+1} f_{n-1} - 2 f_n \bg_n +
      f_{n+1} \bg_{n-1}
      \label{blsymmt} \\
   f_{n+1} f_{n-1} - f_n^2 =  h^2 ||\bg_n||^2 \label{blsymmn}
\end{gather}
\end{subequations}

A two-soliton of the symmetric system is given by the following
solution of the bilinear equations (\ref{blsymmt}-\ref{blsymmn}): 

\begin{subequations}
\begin{align}
   f_n       &= 1 + e^{\eta_{1,n} + \eta_{1,n}^*} +
              e^{\eta_{2,n} + \eta_{2,n}^*} +
              |B_1|^2 e^{\eta_{1,n} + \eta_{1,n}^* + \eta_{2,n} + \eta_{2,n}^*}
              \label{fsymm}\\
   g_n^{(1)} &= \frac{1}{h}
             \left( e^{h p_1} - e^{- h p_1^*} \right) e^{\eta_{1,n}}
             \left( 1 + B_1  e^{\eta_{2,n} + \eta_{2,n}^*} \right)
             \\
   g_n^{(2)} &= \frac{1}{h}
             \left( e^{h p_2} - e^{- h p_2^*} \right) e^{\eta_{2,n}}
             \left( 1 + B_2  e^{\eta_{1,n} + \eta_{1,n}^*} \right)
             \label{gtwosymm}
\end{align}
\end{subequations}

\noindent where $g_n^{(k)}$ is the $k$-th component of $\bg_n$ and

\begin{equation}
  \eta_{j,n} = p_j n h +\frac{i}{h^2} \left(2 - e^{h p_j} -
    e^{- h p_j} \right) t
  \label{deta}.
\end{equation}

\noindent The coefficients $B_j$ are

\begin{equation*}
  B_1 = \dfrac{(e^{h p_1}-e^{h p_2})(e^{h p_1}+e^{h p_2^*})}
              { (e^{h (p_1 + p_2)}+1)(e^{h (p_1 + p_2^*)}-1)}, \qquad
  B_2 = -\dfrac{(e^{h p_1}-e^{h p_2})(e^{h p_1^*}+e^{h p_2})}
               { (e^{h (p_1 + p_2)}+1)(e^{h (p_1^* + p_2)}-1)}
\end{equation*}

\noindent and the complex constants $p_j = a_j - i b_j$ (where $a_j > 0$ and
$j=1,2$) determine the amplitudes and envelope speeds of the solitons. 

In order to see that the above solution (\ref{fsymm}-\ref{gtwosymm}) indeed
gives a two-soliton solution, consider the long-time limits
(i.e. $t \rightarrow \pm \infty$). In the forward ($t \rightarrow +\infty$) and
backward ($t \rightarrow -\infty$) long-time limits, this solution
(\ref{fsymm}-\ref{gtwosymm}) asymptotically approaches the form

\begin{equation*}
 \bq_n^{\pm} \sim \bq_{1,n}^{\pm} + \bq_{2,n}^{\pm}
\end{equation*}

\noindent where $\bq_{j,n}^{\pm} = \bc_{j}^{\pm} \hq_{j,n}^{\pm}$ and $j=1,2$.
The vectors $\bc_j^{\pm}$ are unit vectors  and the scalars $\hq_{j,n}$ are
solitary waves of the form (\ref{discsol}) with $a=a_j$, $b=b_j$
and $\theta=\theta_j^{\pm}$ where $j=1,2$. 
If, without loss of generality, we assume that $a_j$ and $b_j$ are such that
$\sinh(a_1 h) \sin(b_1 h) > \sinh(a_2 h) \sin(b_2 h)$ then

\begin{equation*}
  \theta_1^- =  0, \quad \theta_1^+ = -\log |B_1|, \quad
  \theta_2^- = -\log |B_2|, \quad \theta_2^+ = 0
\end{equation*}

\noindent and

\begin{equation*}
  \bc_1^{\pm} = \left( e^{i \phi_1^{\pm}},  0 \right), \qquad
  \bc_2^{\pm} = \left( 0, e^{i \phi_2^{\pm}} \right)
\end{equation*}

\noindent where

\begin{equation*}
  \phi_1^- =  h b_1, \quad \phi_1^+ = h b_1 + \arg B_1, \quad
  \phi_2^- =  h b_2 + \arg B_2, \quad \phi_1^+ = h b_2.
\end{equation*}

The two-soliton solution (\ref{fsymm}-\ref{gtwosymm}) does not encompass 
the most general case because

\begin{subequations}
\begin{equation}
  \bc_1^- \cdot \bc_2^- = 0  \label{orthogonal:minus}
\end{equation}

\noindent and

\begin{equation}
  \bc_1^+ \cdot \bc_2^+ = 0. \label{orthogonal:plus}
\end{equation}
\end{subequations}

\noindent This solution can be multiplied by a unitary matrix
(the discrete analog of the transformation in eq. \ref{unitrans}) to obtain a
two-soliton solution with any pair of vectors such that
(\ref{orthogonal:minus}) holds (under such a transformation,
(\ref{orthogonal:plus}) will still hold).
Therefore two-soliton solutions given by (\ref{fsymm}-\ref{gtwosymm}) are a
restricted class of two-soliton solutions
(where eq. (\ref{orthogonal:minus}-\ref{orthogonal:plus}) constitutes the
restriction).  

For more than two solitons, Hirota's method can still be applied. 
The $M$-soliton solutions of the symmetric system take on a considerably
more complicated form; they can be
represented as ratios of Pfaffians \cite{Oh98}. Sill, for these
$M$-soliton solutions, either

\begin{subequations}
\begin{equation}
  |\bc_j^- \cdot \bc_k^-| = 1 \label{constraint:parallel}
\end{equation}

\noindent or

\begin{equation}
  |\bc_j^- \cdot \bc_k^-| = 0 \label{constraint:orthogonal}
\end{equation}
\end{subequations}

\noindent and  $|\bc_j^- \cdot \bc_j^+| = 1$ for $j,k=1 \ldots M$
\cite{Oh98,AbTr98}.

Even though analytical formulae exist for
multi-soliton solutions of the symmetric system (with any number, $M$, of
solitons), these solutions are constrained to satisfy
(\ref{constraint:parallel}-\ref{constraint:orthogonal}). Such solutions
are more general, but are like the special class of soliton solutions of the
PDE discussed at the end of Section \ref{pde}. They satisfy the restriction
(\ref{orthogonal:minus}-\ref{orthogonal:plus}).

\subsection{General Soliton Interactions by Numerical Simulation}

Recall that, in the PDE (\ref{vnls}), the polarizations of the individual
solitons in the  multi-soliton solutions (derived by IST) are not
constrained-- i.e there are vector multi-soliton solutions of the PDE
for which the polarizations satisfy $0 < |\bc_j^- \cdot \bc_k^-| < 1$ for
some $j,k$. For the discrete symmetric
system (\ref{sydvnls}), we investigated the general solitary-wave
interaction-- i.e $0 < |\bc_j^- \cdot \bc_k^-| < 1$ 
by numerical simulation because, to date, there are no analytical formulae
covering this general case. Moreover, it is important to determine
whether eq. (\ref{sydvnls}) is integrable. The existence of general soliton
solutions would strongly support the possibility that (\ref{sydvnls}) is
indeed integrable. In addition, as we shall show, the measurement of soliton
parameters in discrete systems requires a novel approach. 

\begin{figure}[p]
  \centerline{
    \psfig{figure=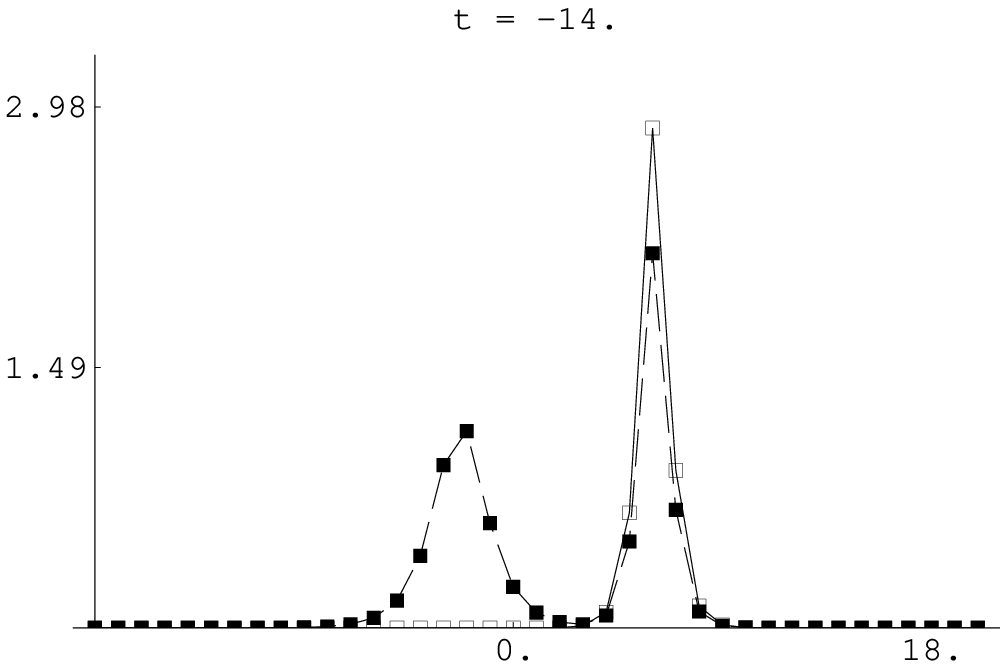,width=3in} \hfill
    \psfig{figure=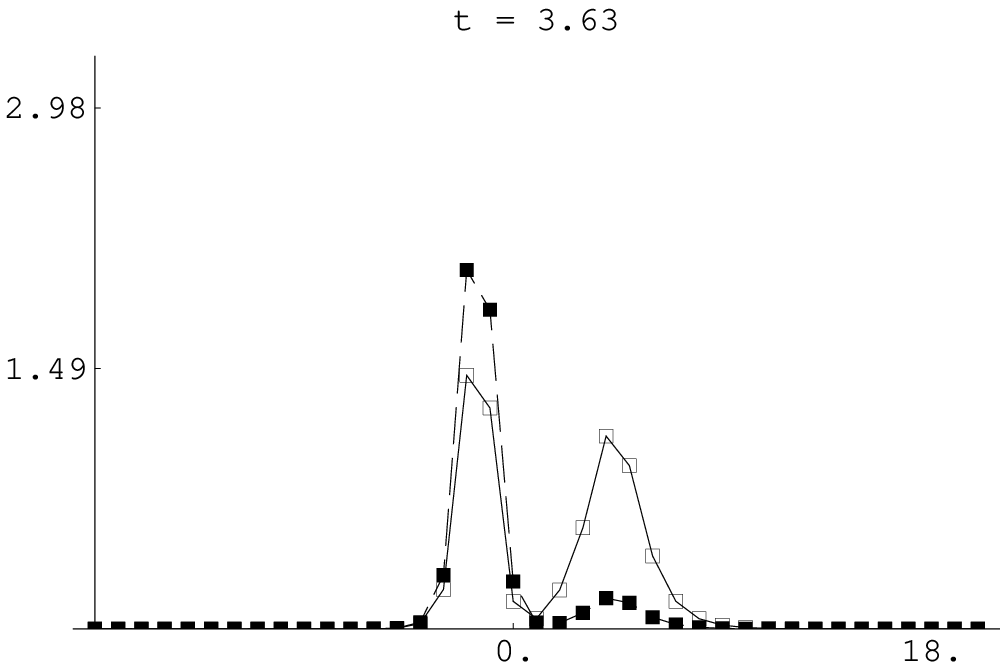,width=3in}} 
   \centerline{
     \psfig{figure=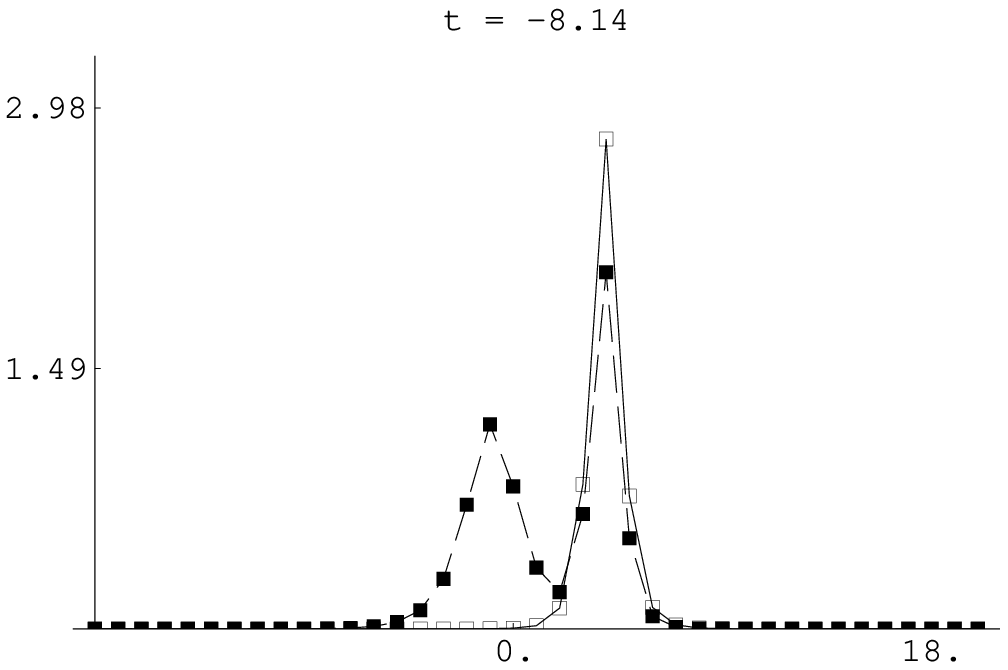,width=3in} \hfill
     \psfig{figure=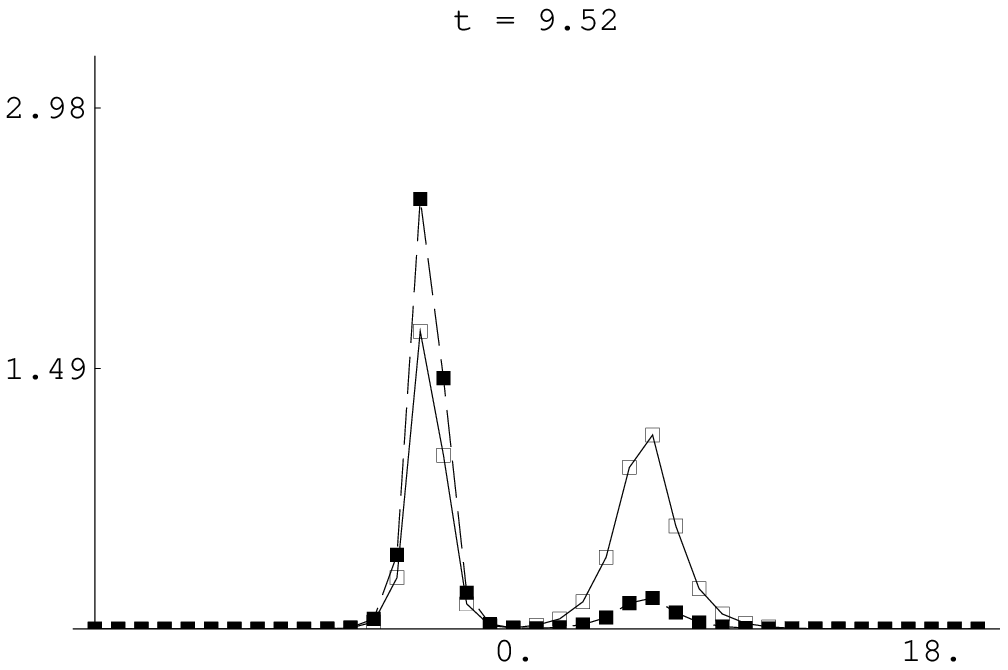,width=3in}} 
   \centerline{
     \psfig{figure=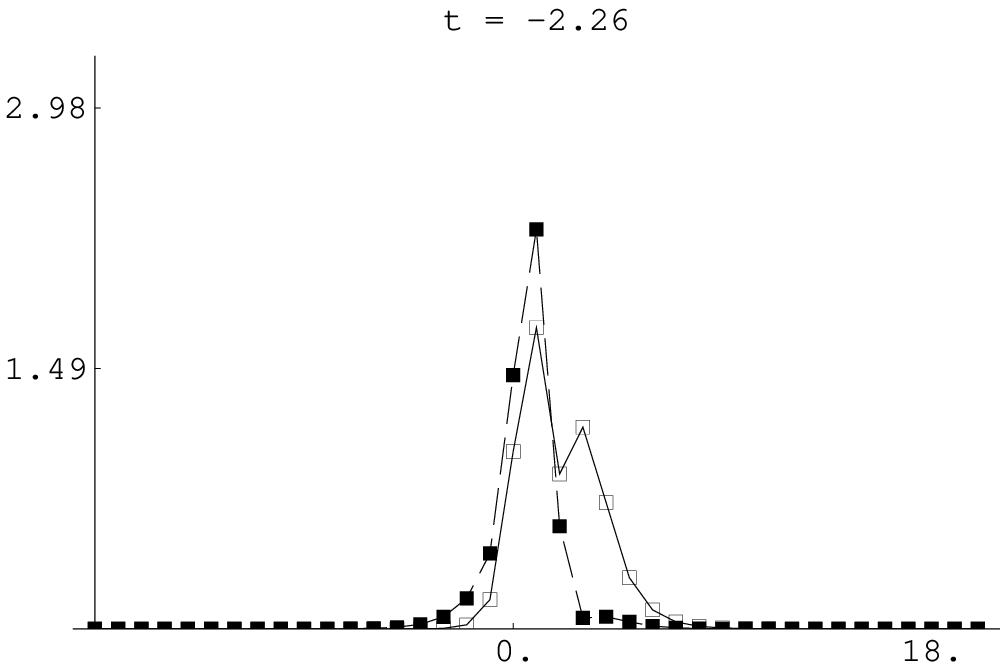,width=3in} \hfill
     \psfig{figure=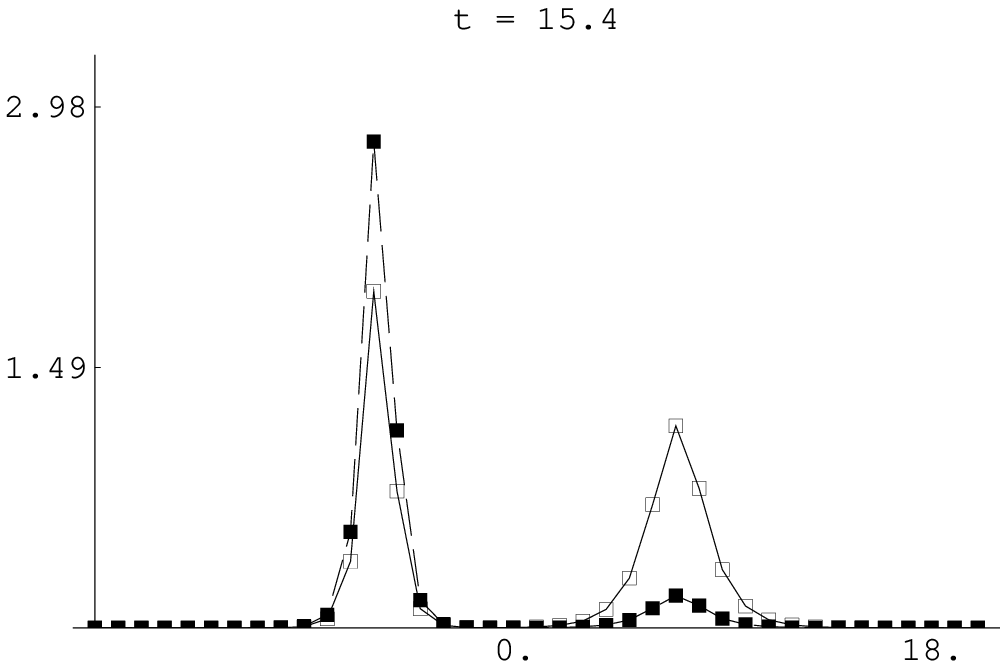,width=3in}} 
   \caption{
     Two-soliton interaction for the symmetric system.
     The filled boxes are $|q_n^{(1)}|$ and the open boxes are
     $|q_n^{(2)}|$. Increasing time is read down column-wise.
     Soliton 1 is on the left and Soliton 2 is on the right
     at $t=-14$. They are reversed at $t=15.4$.
     The soliton parameters are: $a_1 = 1$, $a_2 = 2$,
     $b_1 = .1$, $b_2 = - .1$.
     The polarization vectors are: $\bc_1^- = (1, 0)$,
     $\bc_2^- = (.6 e^{i \pi/3}, .8)$ , at $t=-14$, and 
     $\bc_1^+ = ( .16 e^{i .58 \pi}, .98 e^{-i .16\pi})$,
     $\bc_2^+ = ( .82 e^{i .27 \pi}, .57 e^{-i .03 \pi})$
     at $t=15.4$. This is a typical two-soliton interaction where
     $0 < |\bc_1^- \cdot \bc_2^-| = .6 < 1$,
     $|\bc_1^- \cdot \bc_1^+ | = .16 < 1$ and
     $|\bc_2^- \cdot \bc_2^+ | = .95 < 1$.
   }
   \label{fig:twosym}
\end{figure}

In these simulations, the initial condition was composed
of vector solitary waves $\bq_{j,n}^-$ of the form
$\bq_{j,n}^- = \bc_j^- \hq_{j,n}^-$ where the scalar $\hq_{j,n}^-$ is of the
form (\ref{discsol}). In the initial condition, these vector solitary waves
were well-separated. Then, the symmetric system was integrated in time by an
adaptive Runge-Kutta-Merson routine (from the NAG library) until the peaks
were again well-separated (see Figure \ref{fig:twosym}).
The separation of peaks in the  initial and final conditions makes these
finite-time conditions comparable to, respectively, the backwards and
forwards long-time limits. In the example Figure \ref{fig:twosym}, and in other
simulations, the solitary waves appear to interact without any any radiation
or loss. We confirmed this visual result with striking quantitative
measurements: we measured the difference between the solitary waves resulting
from the numerical simulation and the exact solitary
wave form,

\begin{equation}
  \bq_{j,n}^+ = \bc_j^+ \hq_{j,n}^+ \label{solplus}
\end{equation}

\noindent where $\hq_{j,n}^+$ is of the form (\ref{discsol}).

In order to compare the final-time data of the numerical
simulation with exact solitary-wave solutions it is necessary to estimate the
polarization, $\bc_j^+$, and the shift in the envelope, $\theta_j^+$ for
the  $j$-th solitary wave at the final time
(the values of $\bc_j^-$, and $\theta_j^-$ are fixed by the choice of
initial data). If the $j$-th solitary wave is actually of the form
(\ref{solplus}) at the final time, then the
$\ell$-th component of the vector $\bc_j^+$ satisfies

\begin{equation}
  c_j^{(\ell)+} = e^{ i (b_j h n - \omega_j t_f)}
    \dfrac{q_{j,n}^{(\ell)+}}{\sqrt{|q_{j,n}^{(1)+}|^2 +|q_{j,n}^{(2)+}|^2}}
  \label{estpol}
\end{equation} 

\noindent for all $n$ where $|| \bq_{j,n}^+ ||$ is not
negligible (we denote this set of points $n$ containing the $j$-th
solitary wave by $\Omega_j$). In practice, we compute the estimate $\btc_j^+$
by

\begin{equation*}
  \tc_j^{(\ell)^+}  =
    \text{avg}_{n \in \Omega_j} \quad
    \dfrac{e^{ i (b_j h n - \omega_j t_f)} q_{n}^{(\ell),f}}
      {\sqrt{|q_{n}^{(1),f}|^2 +|q_{n}^{(2),f}|^2}},
\end{equation*}

\noindent where $\ell=1,2$ and $(q_n^{(1),f}, q_n^{(2),f})$ is the
numerical data at the final time, $t=t_f$.
Similarly, if the absolute value of $j$-th solitary wave is a $\sech$
envelope as in (\ref{discsol}), then for all $n \in \Omega_j$, 

\begin{equation}
  \theta_j^+ = a h n - v_{j} t_f -
  \sech^{-1} \left(\frac{|q_{j,n}^{(1)+}|^2 +|q_{j,n}^{(2)+}|^2}{A_j} \right)
  \label{estphase}
\end{equation}

\noindent where $A_j = \tfrac{\sinh (a_j h)}{h}$. Again, we
construct the estimate $\ttheta_j^+$ by taking the average of the
right-hand side of (\ref{estphase}) over $n \in \Omega_j$ where we
substitute $(q_{j,n}^{(\ell)+}, q_{j,n}^{(\ell)+})$ with
$(q_n^{(1),f}, q_n^{(2),f})$, the numerical data at the final time.
Note that we assume that $a_j$ and $b_j$ do not change in the evolution and
hence we do not estimate them in the forward long-time limit.

The difference between the numerically-derived solitary wave
$\bq_{j,n}^+$ and a solitary wave-form of the type
(\ref{solplus}) is measured by
\begin{equation}
  \Delta_j = \max_{n \in \Omega_j}
    \dfrac{\left\| \bq_{j,n}^f - \btq_{j,n}^+ \right\|}{A_j}
  \label{error}
\end{equation} 
where $\bq_{j,n}^f = (q_n^{(1),f}, q_n^{(2),f})$ for
$n \in \Omega_j$ and $\btq_{j,n}^+$ is the estimated soliton wave-form
\begin{equation*}
  \btq_{j,n}^+ = \btc_j^+ A_j e^{-i(b_j h n - \omega_j t_f)}
    \sech \left( a_j h n - v_j t_f - \ttheta_{0,j}^+  \right).  
\end{equation*} 

In order to study the symmetric case in a parameter regime that is
``far'' from the continuum limit, we ran the simulations with amplitude
parameters, $a_j$, and grid sizes $h$ such that the soliton width was
comparable to the grid size (equivalently, $a_j h \approx 1)$. For the
parameters $b_j$ we considered two regimes: (i) $b_j$ such that the spatial
frequency of the complex carrier-wave $e^{i b_j n h}$ was comparable to the
grid size ($b_j h \approx 1$) and (ii) $b_j$ such that the spatial frequency
was large compared to the grid size ($b_j h \approx .1$).  We ran simulations
with several values of $a_j$ for each $b_j$. In particular,
when $b_j h \approx .1$ the speed of the solitary-wave envelopes was slow,
the solitary waves interacted over a long time.

The time-integration routine we used (routine D02BAF from the NAG library)
is adaptive and computes until an internally-defined error tolerance reaches
a user-specified value. In the simulations, when the error tolerance in the
adaptive integration scheme was decreased,
the difference between the data at the final time and a soliton
waveform (\ref{error}) decreased proportionally.
Table \ref{converge} shows data from an
example experiment. In this experiment, the initial data is generic in that 
$0 <| \bc_1^- \cdot \bc_2^- | = .6 < 1$ and, to date,
there is no known analytical formula for the solution. The error,
$\Delta_j$, is proportional to the
error tolerance of the integration. This indicates that
the numerically computed solution converges to an elastic soliton
interaction as the time integration is computed more accurately.
We conclude from numerous  such experiments that, in fact, the vector solitary
waves of the form $\bq_{j,n}^- = \bc^- q_{j,n}^-$ (where the
scalar $q_{j,n}^-$ is of the form of eq. \ref{discsol}) interact
elastically-- i. e. the vector solitary waves behave as true solitons --in the
symmetric system (\ref{sydvnls}). 

For a generic interaction of two solitary waves of the symmetric system,
$0< |\bc_1^- \cdot \bc_2^-| < 1$. In the PDE (\ref{vnls}), such generic
vector soliton interactions result in the shift of the
polarizations of the individual vector
solitons-- i.e. $|\bc_j^- \cdot \bc_j^+| < 1$, $j=1,2$. We observe the same
distinctive vector behavior in numerical simulations of the discrete
symmetric system.  Table \ref{tab:polshift} shows the results of a number
of typical experiments. In a two soliton-interaction we can always pick a
basis for the
polarization such that $\bc_1^- = (1, 0)$ and $\arg c_2^{(2)-} = 0$. Because
$\bc_2^-$ is a unit vector with two components, the absolute value of
$c_2^{(2)-}$ satisfies:
\begin{equation*}
  |c_2^{(2)-}| = \sqrt{1-|c_2^{(1)-}|^2}.
\end{equation*}
Hence, as in Table \ref{tab:polshift},
in order to consider a general two-soliton interaction, we need only
vary $c_2^{(1)-}$. In our simulations,
we see the polarizations of the individual solitary waves shift due to
interaction while the solitary waves retain their shape. In the example
interactions tabulated in Table \ref{tab:polshift}, there are initial
conditions that result in large polarization shifts, that is the inner
products of the forward and backward polarization of an individual vector
solitary wave is small (e.g. $|\bc_1^- \cdot \bc_1^+| = .158$).
This is the intrinsically vector-type
soliton interaction of the solitary waves. Therefore, we conclude from the
numerical simulations that the symmetric system displays general
vector-soliton behavior analogous to VNLS, the PDE continuum limit.

We also considered the following generalization of the two-component
symmetric system
\begin{subequations}
\begin{align}
  i \dtot{t} q^{(1)}_n &=
    \frac{1}{h^2} (q^{(1)}_{n-1} - 2 q^{(1)}_n + q^{(1)}_{n+1}) +
    \left( |q^{(1)}_n|^2 + B |q^{(2)}_n|^2 \right)
      (q^{(1)}_{n-1} + q^{(1)}_{n+1})  \label{xpm1:dsc}\\
  i \dtot{t} q^{(2)}_n &=
    \frac{1}{h^2} (q^{(2)}_{n-1} - 2 q^{(2)}_n + q^{(2)}_{n+1}) +
    \left( B |q^{(1)}_n|^2 + |q^{(2)}_n|^2 \right)
      (q^{(2)}_{n-1} + q^{(2)}_{n+1}). \label{xpm2:dsc}
\end{align}
\end{subequations}
In the case $|B| \neq 1$, we found very strong inelastic effects.
In particular, depending on the initial soliton parameters,
numerical simulations of soliton interaction, the solitons
(i) produced large radiative tails and/or (ii) reflected from one another.
These numerical results suggest that the general elastic interactions
observed in the symmetric system \eqref{sydvnls} are a manifestation of the
integrability of that system. We mention that (\ref{xpm1:dsc}-\ref{xpm2:dsc})
is a an example of the interesting generalization of \eqref{sydvnls} where
the nonlinear term for the $j$-th component,
 $\left( \sum_{k=1}^N |q_n^{(k)}|^2 \right) (q^{(j)}_{n-1} + q^{(j)}_{n+1})$,
is replaced by
 $\left( \sum_{k=1}^N B_{j,k} |q_n^{(k)}|^2 \right)
(q^{(j)}_{n-1} + q^{(j)}_{n+1})$. We expect that this model
will exhibit a range of different inelatic effects between solitary waves
such as those observed in (\ref{xpm1:dsc}-\ref{xpm2:dsc}).
 
The mechanism of the elastic soliton interactions observed for the
symmetric system remains to be completely explained
analytically. More generally, a proof that the symmetric system
can be completely integrated remains as an important open problem.

\section*{Acknowledgments}

This work was supported in part by NSF grants DMS-9404265 and DMS-9703850
as well as U. S. Air Force AASERT grant DOD AF-F49620-93-I-0574 and U. S. Navy
AASERT grant DOD N00014-94-1-0915.

\bibliographystyle{plain}
\bibliography{vnls}

\begin{thebibliography}{10}

\bibitem{AbCh95}
M.~J. Ablowitz, S.~Chakravarty, and B.~M. Herbst.
\newblock Integrability, computation and applications.
\newblock {\em Acta Applicandae Mathematicae}, 39:5--37, 1995.

\bibitem{AbCl91}
M.~J. Ablowitz and P.~A. Clarkson.
\newblock {\em Solitons Nonlinear Evolution Equations and Inverse Scattering}.
\newblock Number 149 in London Mathematical Society Lecture Note Series.
  Cambridge University Press, 1991.

\bibitem{AbHe96}
M.~J. Ablowitz, B.~M. Herbst, and C.~M. Schober.
\newblock Computational chaos in the nonlinear {S}chr{\"o}dinger equation
  without homoclinic crossings.
\newblock {\em Physica A}, 228:212--235, 1996.

\bibitem{AbLa76b}
M.~J. Ablowitz and J.~F. Ladik.
\newblock A nonlinear difference scheme and inverse scattering.
\newblock {\em Studies in Applied Mathematics}, 55:213--229, 1976.

\bibitem{AbTr98}
M.~J. Ablowitz, Y.~Ohta, and A.~D. Trubatch.
\newblock On discretizations of the vector nonlinear {S}chr{\"o}dinger
  equation.
\newblock {A}{P}{P}{M} Preprint 349, University of Colorado-Boulder, March
  1998.

\bibitem{AbSc94}
M.~J. Ablowitz and C.~M. Schober.
\newblock Effective chaos in the nonlinear schr{\"o}dinger equation.
\newblock {\em Contemporary Mathematics}, 172:253--268, 1994.

\bibitem{AbSc93}
M.~J. Ablowitz, C.~M. Schober, and B.~M. Herbst.
\newblock Numerical chaos, roundoff errors and homoclinic manifolds.
\newblock {\em Physical Review Letters}, 71(17):2683--2686, October 1993.

\bibitem{AbSe81}
M.~J. Ablowitz and H.~Segur.
\newblock {\em Solitons and the Inverse Scattering Transfom}.
\newblock Number~4 in SIAM Studies in Applied Mathematics. SIAM, 1981.

\bibitem{CaBi95}
D.~Cai, A.~R. Bishop, N.~Gr{\o}nbech-Jensen, and M.~Salerno.
\newblock Electric field-induced nonlinear bloch oscillations and dynamical
  localization.
\newblock {\em Physical Review Letters}, 74:1186, 1995.

\bibitem{EiLo85}
J.~C. Eilbeck, P.~S. Lombdahl, and A.~C. Scott.
\newblock The discrete self-trapping equation.
\newblock {\em Phyisica}, 16 D:318--338, 1985.

\bibitem{EvMo92}
S.~G. Evangelides, L.~F. Mollenauer, J.~P. Gordon, and N.~S. Bergano.
\newblock Polarization multiplexing with solitons.
\newblock {\em Journal of Lightwave Technology}, 10(1):28--35, January 1992.

\bibitem{HeAb93}
B.~M. Herbst and M.~J. Ablowitz.
\newblock Numerical chaos, symplectic integrators and exponentially small
  splitting distances.
\newblock {\em Journal of Computational Physics}, 105:122--132, 1993.

\bibitem{KeCa86}
V.~M. Kenkre and D.~K. Campbell.
\newblock Self-trapping on a dimer: time-dependent solutions of a discrete
  nonlinear {S}chr{\"o}dinger equation.
\newblock {\em Physical Review B}, 34:4959--4961, 1986.

\bibitem{KeTs87}
V.~M. Kenkre and G.~P. Tsironis.
\newblock Nonlinear effects in quasilinear neutron scattering: exact line
  calculation for a dimer.
\newblock {\em Physical Review B}, 35:1473--1484, 1987.

\bibitem{LaKa97}
T.~I. Lakoba and D.~J. Kaup.
\newblock Perturbation theory for the manakov soliton and its applications to
  pulse propagation in randomly birefringent fibers.
\newblock {\em Physical Review E}, 56(5):6147--6165, November 1997.

\bibitem{Ma74}
S.~V. Manakov.
\newblock On the theory of two-dimensional stationary self-focusing of
  electromagnetic waves.
\newblock {\em Soviet Physics JETP}, 38(2):248--253, 1974.

\bibitem{McOv95}
D.~W. McLaughlin and E.~A.~Overman II.
\newblock {\em Surveys in Applied Mathematics}, volume~1, chapter Whiskered
  Tori for Integrable Pde's: Chaotic behavior in near Integrable Pde's.
\newblock Plenum Press, New York, 1995.

\bibitem{Me87}
C.~R. Menyuk.
\newblock Nonlinear pulse propagation in birefringent optical fibers.
\newblock {\em IEEE Journal of Quantum Electronics}, QE-23(2):174--176,
  February 1987.

\bibitem{Me89}
C.~R. Menyuk.
\newblock Pulse propagation in an elliptically birefringent {K}err medium.
\newblock {\em IEEE Journal of Quantum Electronics}, 25(12):2674--2682,
  December 1989.

\bibitem{Oh98}
Y.~Ohta.
\newblock Pfaffian solutions for coupled discrete nonlinear schr{\"o}dinger
  equation.
\newblock {\em Chaos, Solitons and Fractals}, to appear.
\newblock Proceedings of Brussels Meeting II: Integrability and Chaos in
  Discrete Systems (Brussels, 2-6 July 1997).

\bibitem{ShDo97}
V.~S. Shchesnovich and E.~V. Doktorov.
\newblock Perturbation theory for solitons of the {M}anakov system.
\newblock {\em Physical Review E}, 55(6):7626--7635, June 1997.

\bibitem{Su94}
Y.~B. Suris.
\newblock A discrete-time {G}arnier system.
\newblock {\em Physics Letters. A}, 189(4):281--289, 1994.

\end{thebibliography}

\begin{table}[p]
  \begin{center}
    \begin{tabular}{|c|c|c|}
      \hline
                 & \multicolumn{2}{c|}{Error: $\log_{10} \Delta_j$} \\
      \hline 
      $tol$      & soliton 1   & soliton 2 \\
      \hline \hline 
      $10^{-6}$  & $-4.11$     & $ -5.97$ \\
      $10^{-7}$  & $-5.33$     & $ -6.99$ \\
      $10^{-8}$  & $-6.58$     & $ -8.00$ \\
      $10^{-9}$  & $-7.84$     & $ -9.01$ \\
      $10^{-10}$ & $-9.12$     & $ -10.0$ \\
      \hline
    \end{tabular}
    \caption{Difference between solution at the final time and
      a soliton wave-form for a two solitary-wave interaction.
      Soliton parameters: $a_1 = 2$, $b_1 = 0.1$,
      $\bc_1^- = (1, \quad 0)$, $a_2 = 1$,
      $b_2 = -0.1$, 
      $\bc_2^- = ( \tfrac{1}{\sqrt{2}} \quad  \tfrac{1}{\sqrt{2}} )$.
      ``$tol$'' is the error tolerance allowed by the NAG routine D02BAF.
      The errors in the table are the $\log_{10}$ of the error given by
      (\ref{error}).
      The error decreases proportionally with the error
      tolerance used in the simulation. This indicates that the errors
      are due to the time integration and that, therefore, the solitary
      waves interact as solitons.
      }
     \label{converge}
  \end{center}
\end{table}

\begin{table}[p]
  \begin{center}
    \begin{tabular}{|c|c|c||c|c|c|}
     \hline
      $\bc_2^-$ & $|\bc_1^- \cdot \bc_1^+|$ & $|\bc_2^- \cdot \bc_2^+|$ &
      $\bc_2^-$ & $|\bc_1^- \cdot \bc_1^+|$ & $|\bc_2^- \cdot \bc_2^+|$ \\     
     \hline
     \hline
     $(0.2 e^{i 0}, .98 e^{i 0})$        & 0.843  & 0.985 &
     $(0.2  e^{i \pi /3}, .98 e^{i 0})$  & 0.843  & 0.985 \\
     $(0.4  e^{i 0}, .92 e^{i 0})$       & 0.465  & 0.958 &
     $(0.4  e^{i \pi /3}, .92 e^{i 0})$  & 0.465  & 0.958 \\
     $(0.6  e^{i 0}, 0.8 e^{i 0})$       & 0.158  & 0.947 &
     $(0.6  e^{i \pi /3}, 0.8 e^{i 0})$  & 0.158  & 0.947 \\
     $(0.8  e^{i 0}, 0.6 e^{i 0})$       & 0.544  & 0.962 & 
     $(0.8  e^{i \pi /3}, 0.6 e^{i 0})$  & 0.544  & 0.962 \\
     \hline
     $(0.2  e^{i \pi /2}, .98 e^{i 0})$  & 0.843 & 0.985 &
     $(0.2  e^{i \pi}, .98 e^{i 0})$     & 0.843 & 0.985 \\
     $(0.4  e^{i \pi /2}, .92 e^{i 0})$  & 0.465 & 0.958 &
     $(0.4  e^{i \pi}, .92 e^{i 0})$     & 0.465 & 0.958 \\
     $(0.6  e^{i \pi /2}, 0.8 e^{i 0})$  & 0.158 & 0.947  &
     $(0.6  e^{i \pi}, 0.8 e^{i 0})$     & 0.158 & 0.947 \\
     $(0.8  e^{i \pi /2}, 0.6 e^{i 0})$  & 0.544 & 0.962 &
     $(0.8  e^{i \pi}, 0.6 e^{i 0})$     & 0.544 & 0.962 \\
     \hline
    \end{tabular}
    \caption{Shift in polarization in the two-soliton interaction of
      the symmetric system. The soliton parameters are $a_1 = 1$,
      $a_2 = 2$, $b_1 = .1$, $b_2 = - .1$. The polarizations
      before interaction are $\bc_1^- = ( 1, 0)$ and $\bc_2^-$ as given
      in the table. The polarizations after interaction are $\bc_1^+$ and
      $\bc_2^+$. The values $|\bc_j^- \cdot \bc_j^+| < 1$ indicate that the
      polarization vectors are shifted by the soliton interaction.
      These results were obtained by numerical simulation.}
    \label{tab:polshift}
  \end{center}
\end{table}
\end{document}